\documentclass[nofootinbib,twocolumn,floatfix]{revtex4}
\usepackage{graphicx}
\usepackage{amsfonts}
\usepackage{amssymb}
\usepackage{amsmath}
\usepackage{latexsym}
\usepackage{subfloat}

\begin{document}

\preprint{ICG preprint 04/xx}

\title{Naked shell singularities on the brane}

\author{Sanjeev S.~Seahra}
\email{sanjeev.seahra@port.ac.uk}%
\affiliation{Institute of Cosmology \& Gravitation, University of
Portsmouth, Portsmouth, PO1 2EG, UK}

\setlength\arraycolsep{2pt}
\newcommand*{\di}{\partial}
\newcommand*{\V}{{\mathcal V}^{(k)}_d}
\newcommand*{\volume}{\sqrt{\sigma^{(k,d)}}}
\newcommand*{\OneTwo}{{(1,2)}}
\newcommand*{\onetwo}{{1,2}}
\newcommand*{\Lm}{{\mathcal L}_m}
\newcommand*{\stm}{{\textsc{stm}}}
\newcommand*{\Hm}{{\mathcal H}_m}
\newcommand*{\hatHm}{\hat{\mathcal H}_m}
\newcommand*{\Ldust}{{\mathcal L}_\mathrm{dust}}
\newcommand*{\maxsym}{{\mathbb S}_d^{(k)}}
\newcommand*{\sn}{{\mathrm{sn}}}
\newcommand*{\cn}{{\mathrm{cn}}}
\newcommand*{\nc}{{\mathrm{nc}}}
\newcommand*{\Jacobisc}{{\mathrm{sc}}}
\newcommand*{\ansatz}{{\emph{ansatz}}}
\newcommand*{\ds}[1]{ds^2_\text{\tiny{($#1$)}}}
\newcommand*{\kret}[1]{\mathfrak{K}_\text{\tiny{($#1$)}}}
\newcommand*{\ads}[1]{{AdS$_{#1}$}}

\date{\today}

\begin{abstract}

By utilizing non-standard slicings of 5-dimensional Schwarzschild
and Schwarzschild-AdS manifolds based on isotropic coordinates, we
generate static and spherically symmetric braneworld spacetimes
containing shell-like naked null singularities.  For planar
slicings, we find that the brane-matter sourcing the solution is a
perfect fluid with an exotic equation of state and a pressure
singularity where the brane crosses the bulk horizon.  From a
relativistic point of view, such a singularity is required to
maintain matter infinitesimally above the surface of a black hole.
From the point of view of the AdS/CFT conjecture, the singular
horizon can be seen as one possible quantum correction to a
classical black hole geometry.  Various generalizations of planar
slicings are also considered for a Ricci-flat bulk, and we find
that singular horizons and exotic matter distributions are common
features.

\end{abstract}

\maketitle

\section{Introduction}\label{sec:intro}

Recent advances in non-perturbative string theory have raised the
prospect that our universe is a 4-dimensional hypersurface (brane)
embedded within some higher-dimensional manifold with large extra
dimensions.  A phenomenological 5-dimensional realization of this
idea was proposed by Randall \& Sundrum (RS) in 1999
\cite{Ran99a}, which involved one or two 4-dimensional Minkowski
branes embedded in an anti-deSitter `bulk' 5-manifold (\ads{5}).
One of the most attractive features of this `braneworld' model is
the fact that the 5-dimensional graviton zero mode is sharply
confined near the `visible brane' representing our universe,
implying that the force of gravity has the appropriate Newtonian
behaviour at large distances. This automatically makes the
one-brane model in excellent agreement with most astrophysical
tests of general relativity in the weak gravity regime.

But this virtue is also somewhat of a detriment, because we must
turn to strong gravity phenomena in order to test the model, and
thereby the stringy ideas that motivated it.  The appropriate
formalism to deal with non-trivial curvature in the braneworld was
developed by Shiromizu et al.~\cite{Shiromizu:1999wj} shortly
after the RS model first appeared.  They obtained an effective
4-dimensional Einstein equation that was in part sourced by the
(traceless) projection $\mathcal{E}_{\mu\nu}$ of the bulk Weyl
tensor onto the brane. But this tensor did not come with a
brane-based equation of motion, which means that the 4-dimensional
effective theory is not closed --- one needs to know about the
geometry of the bulk to fully specify the dynamics of the brane.
If one insists on using a purely brane-based formalism, the
precise form of $\mathcal{E}_{\mu\nu}$ is somewhat arbitrary.

It turns out this ambiguity is not a big problem for braneworld
cosmology.  If one has a cosmological brane which retains a
Friedmann-Robertson-Walker (FRW) form for all time, it follows
that the bulk spacetime shares the same symmetries; i.e, the bulk
is the product of $\mathbb{R}^2$ with a maximally symmetric
3-space, and is sourced by a negative cosmological constant. Under
such circumstances, the 5-dimensional version of Birkhoff's
theorem states that the bulk is necessarily isometric to the
5-dimensional Schwarzschild-anti-deSitter (S-\ads{5}) solution.
This forces $\mathcal{E}_{\mu\nu}$ to take the form of the
stress-energy tensor of a cosmological radiation field whose
amplitude is controlled by the mass of the bulk black hole.

But there is another strong gravity phenomenon that is at least as
important as cosmology, namely black holes.  Spherically-symmetric
black hole 4-metrics have fewer symmetries than their FRW
counterparts, which implies that the bulk geometry is not nearly
as constrained as it is for braneworld cosmology.  In turn, this
means that $\mathcal{E}_{\mu\nu}$ is undetermined by simply
specifying that the brane is spherically symmetric and devoid of
matter.  Stated in another way, there is no 4-dimensional Birkhoff
uniqueness theorem for braneworld black holes; the bulk Weyl
contribution acts as an arbitrary effective source.

Hence, there are many possible candidates for the `right' model of
a braneworld black hole.  One way to get at them is to set the
matter content of the brane to zero and fine tune its tension,
which makes the effective brane field equation ${}^{(4)}R_{\mu\nu}
= -\mathcal{E}_{\mu\nu}$.  This has been solved under spherically
symmetric conditions by a number of authors, but they all had to
assume something about the form of $\mathcal{E}_{\mu\nu}$.  For
example, there is the tidal Reissner-Nordstr\"om solution of
Dadhich et al.~\cite{Dadhich:2000am}, or the line elements of
Gregory et al.~\cite{Gregory:2004vt} that assume an equation of
state for the `Weyl fluid.'  A different line of attack comes from
trying to solve the (scalar) field equation ${}^{(4)}R = $
constant \cite{Bronnikov:2002rn,Bronnikov:2003gx}, which comes
from the contracted Gauss-Codazzi equations.  Recently, the
so-called `gradient-expansion' method has been applied to problem
in an effort to systematically include effects of the extra
dimension on the brane metric \cite{McFadden:2004ni}.  Several
workers have also looked at dynamical case of gravitational
collapse on the brane, and have come to the conclusion that the
exterior brane metric to a collapsing star cannot be static
\cite{Bruni:2001fd} and in some cases is not even a vacuum
\cite{Casadio:2004nz}. The cumulative effect of these efforts has
been to create a veritable zoo of black hole candidates, some of
which have reasonable physical properties.

But all of these models are somewhat unsatisfactory because of
ignorance of the bulk geometry.  One does not know if any
singularities present extend off the brane, or the nature and
shape of any 5-dimensional horizons.  Hence, one cannot study the
thermodynamics of such objects.  Perturbations of these geometries
are also ill-defined because of the under-determined nature of the
effective theory. This means that we cannot address the stability
of these models nor their gravity wave signatures, which may be an
important observational test of extra dimensions
\cite{Seahra:2004fg}.  It is in theory possible to obtain the bulk
geometry by evolving the 4-metric off the brane, but the problem
is analytically complicated \cite{Kanti:2001cj,Casadio:2002uv} and
robust numerical progress can only be made for `small' black holes
\cite{Chamblin:2000ra,Wiseman:2001xt,Kudoh:2003xz}.

However, there is at least one credible alternative to these
brane-based approaches.  Instead of trying to deal with effective
field equations, one can take known 5-dimensional solutions and
identify branes as slices embedded therein.  The most successful
example of this procedure is actually 4-dimensional. Emparan et
al.~\cite{Emparan:1999wa} considered a simple slicing of the
4-dimensional C-metric.  The slice had an extrinsic curvature
proportional to its induced metric, implying a pure tension brane,
and a $(2+1)$-dimensional black hole intrinsic geometry.
Furthermore, the bulk was entirely regular. But unfortunately,
there currently is no 5-dimensional generalization of the C-metric
that allows the same construction for a $(3+1)$-dimensional brane
black hole, despite concerted efforts to find one
\cite{Charmousis:2003wm}.  A different possibility for the bulk
manifold is the 5-dimensional black string solution, which is  a
simple warped-product model where the brane metric is precisely
Schwarzschild \cite{Chamblin:1999by,Nojiri:2000yr}. Unfortunately
in a one brane model, this metric is subject to the well-known
long-wavelength Gregory-Laflamme (GL) instability
\cite{Gregory:2000gf}; however, one can engineer a two-brane
scenario where the GL instability is cut-off \cite{Seahra:2004fg}.

But why has it been so hard to find a brane localized black hole
solution in 5 dimensions?  Separately, Tanaka \cite{Tanaka:2002rb}
and Emparan et al.~\cite{Emparan:2002px} have conjectured that the
reason has to do with the AdS/CFT correspondence
\cite[refs.~therein]{Aharony:1999ti}, which states that the
dynamics of an \ads{n} manifold are formally dual to behaviour of
an $(n-1)$-dimensional conformal field theory (CFT) living on its
boundary.  The authors noted that in the 4-dimensional model of
ref.~\cite{Emparan:1999wa}, the $\mathcal{E}_{\mu\nu}$ part of the
brane's effective stress-energy tensor took the form a
quantum-corrected $(2+1)$-black hole. That is, the solution on the
boundary of \ads{4} was derived from the backreaction of a quantum
field on the classical lower-dimensional black hole geometry.
Extending the logic to one dimension higher, we are led to believe
that the $(3+1)$-braneworld black hole ought to take the form of
Schwarzschild subject to quantum corrections
\cite{Balbinot:1999vg}. The precise form of the correction depends
on the choice made for the quantum vacuum. One possibility has the
black hole radiating its mass away via the Hawking effect (which
is what is conventionally regarded as the end-state of
gravitational collapse on the brane) another involves the black
hole in thermal equilibrium with a heat bath at infinity. Yet
another choice yields a static configuration with a singularity
where the horizon used to be. We now see the difficulty in finding
the 4-dimensional brane black hole; all of these possibilities
represent significant departures from the canonical Schwarzschild
geometry.

The purpose of this paper is to develop spherically symmetric and
static braneworld models using methods inspired from the
successful construction of 2-brane localized black holes.  In
particular, we will be considering various slicings of
5-dimensional black hole metrics, both with and without a negative
cosmological constant $\Lambda_5 = -6/\ell^2$.  We work in
isotropic coordinates, which are developed in
Sec.~\ref{sec:isotropic}.  In Sec.~\ref{sec:branes from isotropic
charts}, we study the simplest possible braneworlds based on a
planar slicing of the 5-manifold through the event horizon.  The
basic methodology is similar to the 4-dimensional
`displace--cut--reflect' procedure for constructing thin-disk
solutions to the Einstein equations \cite{Vogt:2003hi}.
Intriguingly, we find that brane 4-geometry involves a naked
shell-singularity for all cases we consider; i.e., with $\Lambda_5
\le 0$.  This is as expected from the AdS/CFT considerations
mentioned above.  Because the planar slicing is selected on purely
geometric grounds, the extrinsic curvature and matter content of
the braneworlds is not freely specifiable, it is rather forced
upon us.  We find that the models are supported by a non-trivial
perfect fluid with a pressure singularity where the brane
intersects the 5-dimensional horizon.  Such a singularity could
have been predicted on physical grounds: One requires an infinite
amount of force to keep matter suspended infinitesimally above the
surface of a black hole.  Hence, an infinite pressure gradient is
needed to keep the brane matter static.  Finally, in
Sec.~\ref{sec:non-planar} we consider quite general non-planar
slices.  These include slicings with vanishing Ricci scalar,
radial pressure, tangential pressure, and extrinsic curvature, as
well as slicings with isotropic pressure and pure tension branes.
In the last case, the only solution we find corresponds to an
Einstein static universe coincident with the photon sphere of the
bulk black hole.  Sec.~\ref{sec:conclusions} is reserved for
conclusions and final comments.

\paragraph*{Conventions}  We employ the `mostly positive' metric
signature.  Lowercase Latin indices run from 0 to 4 and lowercase
Greek indices run from 0 to 3.  Metric compatible covariant
derivatives on 5-manifolds are denoted by $\nabla_a$; while on
4-dimensional submanifolds (3-branes) they are denoted by
$\nabla_\alpha$.

\section{Transformations from spherical to isotropic
coordinates}\label{sec:isotropic}

The purpose of this section is to describe how isotropic
coordinates can be constructed for a certain class of spherically
symmetric manifolds in an arbitrary number of dimensions, and to
derive explicit coordinate transformations for two special
5-dimensional cases.  These special cases will be used in the next
section to construct braneworld shell solutions associated with
vacuum and Schwarzschild-AdS bulk manifolds, respectively.

\subsection{General transformations for a class of
$(d+2)$-dimensional spherically symmetric manifolds}

We begin by considering a fairly wide $(d+2)$-dimensional class of
spherically symmetric manifolds $(M,g)$ whose line element can be
expressed as
\begin{equation}\label{general line element}
    \ds{M} = -f(R)\,dt^2 + f^{-1}(R)\,dR^2 + R^2 \, d\Omega_d^2,
\end{equation}
where $d\Omega_d^2$ is the interval on a unit $d$-sphere.  Our
goal is to find a coordinate transformation that puts this in the
isotropic form
\begin{equation}\label{isotropic line element}
    \ds{M} = -H(\rho)\,dt^2 + G(\rho) [d\rho^2 + \rho^2 \,
    d\Omega_d^2].
\end{equation}
This line element is called isotropic because a further simple
coordinate transformation yields
\begin{equation}
    \ds{M} = -H(\rho)\,dt^2 + G(\rho) \sum^{d+1}_{i=1} dx_i^2,
\end{equation}
with $\rho = \sqrt{x_1^2 + x_2^2 + \cdots + x_{d+1}^2}$.  In these
coordinates each of the spatial directions is on the same footing,
hence the moniker ``isotropic.''

It is easy to see that the coordinate transformation from
(\ref{general line element}) to (\ref{isotropic line element})
must satisfy
\begin{equation}
    \left( \frac{dR}{d\rho} \right)^2 = f(R) G(\rho),
    \quad R^2 = G(\rho) \rho^2.
\end{equation}
This set of equations is solved by
\begin{equation}\label{rho integral}
    \rho(R) = \exp \int^R_{R_0} \frac{du}{\sqrt{u^2 f(u)}},
\end{equation}
which must be inverted to obtain $R = R(\rho)$.  Here, $R_0$ is
some fiducial lower limit of integration that enforces $\rho(R_0)
= 1$. Assuming that such an inversion is possible, we have the
following implicit representations of the isotropic metric
functions:
\begin{equation}\label{general metric functions}
    G(\rho) = \frac{R^2(\rho)}{\rho^2}, \quad H(\rho) =
    f(R(\rho)).
\end{equation}
Hence the required coordinate transformation is found.

We make two comments before proceeding:  First, it is
straightforward to confirm that if we adopt the familiar
4-dimensional Schwarzschild solution with $d = 2$ and $f(R) = 1 -
2M/R$, we obtain the usual isotropic coordinate patch found in
standard textbooks. Second, we note that the integral in (\ref{rho
integral}) is complex if $f(R) < 0$ anywhere in the interval
$[R_0,R]$. Therefore, if the line element (\ref{general line
element}) represents a black hole manifold, then the isotropic
coordinate patch can only be used to cover the portion outside the
horizon; i.e., the part of the manifold with $f > 0$. We will
return to this point below.

\subsection{The 5-dimensional Schwarzschild black hole in
isotropic coordinates}\label{sec:iso Schwarzschild}

We now turn our attention to the 5-dimensional Schwarzschild black
hole.  Usually, this is expressed as
\begin{equation}
    \ds{M} = -\left(1 -
\frac{\mathcal{R}_0^2}{\mathcal{R}^2}\right) d\mathcal{T}^2 +
\left(1 -
\frac{\mathcal{R}_0^2}{\mathcal{R}^2}\right)^{-1}d\mathcal{R}^2 +
\mathcal{R}^2 d\Omega_3^2,
\end{equation}
Here, $\mathcal{R}_0$ represents the position of the black hole
horizon and is also related to the ADM mass of the central object.
For our purposes, it is useful to adopt dimensionless radial and
time coordinates by making the changes $\mathcal{R} \rightarrow
\mathcal{R}_0 \times R$ and $\mathcal{T} \rightarrow \mathcal{R}_0
\times t$.  If this is accompanied by a simultaneous scaling of
the interval $\ds{M} \rightarrow \mathcal{R}_0^2 \times \ds{M}$,
we have the line element is in the standard form (\ref{general
line element}) with
\begin{equation}
    f(R) = 1 - 1/R^2, \quad d = 3.
\end{equation}
Notice that when we are working in dimensionless coordinates,
there are no freely specifiable parameters in the solution, and
the horizon is always at $R = 1$.

By application of the formula (\ref{rho integral}) with $R_0$ set
to unity --- as dictated by the horizon position in these
coordinates --- we obtain the transformations
\begin{equation}\label{Schwarzschild transformation}
    \rho = R + \sqrt{R^2 -1}, \quad R = \frac{\rho^2 +1}{2\rho}.
\end{equation}
From these, it is clear that the $\rho$ coordinate is only well
defined for $R > 1$; i.e., outside the black hole horizon.  The
explicit form of the isotropic metric functions is
\begin{subequations}\label{isotropic Schwarzschild}
\begin{eqnarray}
    H(\rho) & = & \left(\frac{\rho^2 - 1}{\rho^2+1}\right)^2, \\
    G(\rho) & = & \left(\frac{\rho^2+ 1}{2\rho^2} \right)^2.
\end{eqnarray}
\end{subequations}
In order to check our work, we have confirmed by direct
calculation of the Einstein tensor that these metric functions
represent a 5-dimensional vacuum solution.  Intriguingly, they
provide a solution for all $\rho$, not just $\rho > 1$ --- this
will be important later on.  It is also interesting to note that
the Killing vector $\di_t$ in these coordinates becomes null at
$\rho = 1$, but is nowhere spacelike. This is a direct affirmation
of our previous conclusion that the isotropic coordinates do not
cover the region inside the horizon, which is characterized by
$\di_t \cdot \di_t > 0$.

\subsection{The 5-dimensional Schwarzschild-AdS black hole in
isotropic coordinates}

Moving on, we come to the case of a 5-dimensional black hole
sourced by a negative cosmological constant; i.e., the
Schwarzschild-AdS 5-manifold (S-AdS$_5$). The conventional form of
such a solution is
\begin{subequations}
\begin{eqnarray}
    \ds{M} & = & -\mathcal{F}\,d\mathcal{T}^2 + \mathcal{F}^{-1}
    \,d\mathcal{R}^2 + \mathcal{R}^2 \, d\Omega_3^2, \\
    \mathcal{F} & = & \mathcal{F}(\mathcal{R}) = 1 -
    \frac{\mathcal{R}_0^2}{\mathcal{R}^2} +
    \frac{\mathcal{R}^2}{\ell^2}.
\end{eqnarray}
\end{subequations}
Here, $\mathcal{R}_0$ is again related to the ADM mass of the
black hole while $\ell$ is related to the (negative) cosmological
constant.  It is convenient to rewrite $\mathcal{F}$ as
\begin{equation}
    \mathcal{F} = \frac{(\mathcal{R}^2 + \mathcal{R}_+^2)(\mathcal{R}^2 -
    \mathcal{R}_-^2)}{\mathcal{R}^2 \ell^2},
\end{equation}
where
\begin{equation}
    \mathcal{R}^2_\pm = \frac{\ell^2}{2} \left( \sqrt{
    \frac{4\mathcal{R}_0^2}{\ell^2} + 1} \pm 1 \right).
\end{equation}
When the solution is written in this way, it is apparent that
there is an event horizon at $\mathcal{R} = \mathcal{R}_-$.  We
again wish to make use of dimensionless coordinates, this time
defined by the substitutions:
\begin{eqnarray}\nonumber
    \mathcal{R} & \rightarrow & R \times \mathcal{R}_-, \\ \mathcal{T}
    & \rightarrow & t \times \mathcal{R}_-, \\ \nonumber \ds{M} &
    \rightarrow & \ds{M} \times \mathcal{R}_-^2.
\end{eqnarray}
We must also define new parameters as follows:
\begin{equation}
    \gamma = \frac{\ell}{\mathcal{R}_-}, \quad a_\gamma \equiv \frac{\mathcal{R}_+}{\mathcal{R}_-}
    = \sqrt{ \gamma^2+1 }.
\end{equation}
With these manipulations, the S-AdS$_5$ line element can be
expressed in the form of equation (\ref{general line element})
with
\begin{equation}\label{S-AdS f}
    f(R) = \frac{(R^2 + a_\gamma^2)(R^2-1)}{\gamma^2 R^2}, \quad
    d = 3,
\end{equation}
and $t$ and $R$ as dimensionless coordinates.  As in the vacuum
example discussed in the previous subsection, the horizon is
located at $R = 1$; but unlike the Schwarzschild case, there is an
adjustable parameter in the dimensionless solution, namely
$\gamma$.

We now obtain the S-AdS$_5$ line element in isotropic coordinates.
The first step is to put the S-AdS$_5$ expression for $f$ into our
general expression for $\rho(R)$; i.e., equation (\ref{rho
integral}).  We again set the lower limit of integration $R_0$ as
the position of the horizon at $R=1$.  This results in
\begin{equation}\label{S-AdS rho}
    \rho(R) = \exp \left[ \frac{1}{\xi_\gamma}
    F\left( \sqrt{1-\frac{1}{R^2}}, s_\gamma \right) \right].
\end{equation}
Here,
\begin{equation}\label{s defn}
    s_\gamma \equiv \sqrt{\frac{
    \gamma^2+1}{\gamma^2+2}}, \quad \xi_\gamma \equiv
    \sqrt{\frac{\gamma^2+2}{\gamma^2}},
\end{equation}
and $F$ is an incomplete elliptic integral of the first kind,
defined by\footnote{The reader should be wary of a common
alternative definition of $F$, namely
\begin{displaymath}
    F(z,m) \equiv \int_0^{\sin z} \frac{dt}{\sqrt{(1-t^2)(1- m t^2)}}.
\end{displaymath}
Our definition (\ref{F definition}) matches the one found in the
\textsc{Maple} symbolic computation software.}
\begin{equation}\label{F definition}
    F(z,k) \equiv \int_0^z \frac{dt}{\sqrt{(1-t^2)(1-k^2 t^2)}}.
\end{equation}
One comment about this coordinate transformation is in order: the
limit of $\rho(R)$ as $R \rightarrow \infty$ is a constant value,
namely
\begin{equation}\label{max rho}
    \rho_\mathrm{max} = \exp\left[ \frac{K ( s_\gamma )}{\xi_\gamma}
     \right],
\end{equation}
where $K(k) = F(1,k)$ is a complete elliptic integral of the first
kind.  So, the $R \rightarrow \rho$ transformation maps the
semi-infinite interval $R \in (1,\infty)$ onto some finite region
$\rho \in (1,\rho_\mathrm{max})$.  This is unlike the vacuum case
above, since equation (\ref{Schwarzschild transformation}) implies
that $\rho \rightarrow 2R$ as $R \rightarrow \infty$.

The above expression (\ref{S-AdS rho}) for $\rho$ as a function of
$R$ is indeed invertible with the aid of the Jacobi $\sn$ and
$\cn$ functions, which are implicitly defined by
\begin{equation}
    \sn(F(z,k),k) = z,
\end{equation}
and
\begin{equation}
    \cn(z,k) \equiv \cos \{ \arcsin [
    \sn(z,k)] \}.
\end{equation}
In many respects, these behave like the familiar trigonometric
sine and cosine functions --- in particular, they are periodic in
their first argument. The old radius $R$ as a function of the
isotropic radius $\rho$ is then given by
\begin{equation}\label{S-AdS R}
    R(\rho) = \nc\left( \varphi(\rho) , s_\gamma \right),
\end{equation}
where $\nc(z,k) = 1/\cn(z,k)$ and we have defined
\begin{equation}
    \varphi(\rho) \equiv \xi_\gamma \ln\rho.
\end{equation}
The periodic nature of the $\nc$ function in $R(\rho)$ means that
we should restrict $\rho$ to lie within some finite interval in
order to have a sensible coordinate transformation --- however,
this is no surprise because we have already determined from
equation (\ref{S-AdS rho}) that $\rho(R) \in
(1,\rho_\mathrm{max})$ for $R \in (1,\infty)$.

Finally, the isotropic metric functions $G$ and $H$ are easily
found from equations (\ref{general metric functions}), (\ref{S-AdS
f}) and (\ref{S-AdS R}):
\begin{subequations}
\begin{eqnarray}
    G(\rho) & = & \frac{ \nc^2\left( \varphi(\rho) , s_\gamma
    \right) }{\rho^2}, \\
    H(\rho) & = & \frac{ \Jacobisc^2 \left(
    \varphi(\rho) , s_\gamma \right) + a_\gamma^2 \sn^2
    \left( \varphi(\rho) , s_\gamma \right)  } {\gamma^2},
\end{eqnarray}
\end{subequations}
where the Jacobi $\Jacobisc$ function is defined like a tangent;
i.e., $\Jacobisc(z,k) \equiv \sn(z,k)/\cn(z,k)$.  We note
$\sn(0,k) = \Jacobisc(0,k) = 0$ for all $k$ and $\varphi(1) = 0$,
therefore $H(1) = 0$.  That is, at $\rho = 1$ the Killing vector
$\di_t$ becomes null.  Elsewhere, $H(\rho)$ is explicitly
non-negative, which again confirms that the isotropic coordinates
only cover the region outside the black hole horizon with $\di_t
\cdot \di_t < 0$.  Again, we have confirmed by direct computation
that the above metric functions solve the 5-dimensional field
equations:
\begin{equation}
    G_{ab} = \frac{6}{\gamma^2} g_{ab},
\end{equation}
for all $\rho$.

\section{Braneworlds from planar slicings of isotropic
charts}\label{sec:branes from isotropic charts}

In the previous section, we developed isotropic coordinate patches
for a fairly wide class of spherically symmetric manifolds and for
two special 5-dimensional cases.  We now attempt to generate
braneworld models from these special cases by considering their
planar slicings, first for the purely Schwarzschild bulk spacetime
and then for the S-AdS$_5$ manifold. While the latter is more
technically complicated than the former, we will see that the
basic physics associated with both cases is remarkably similar.

\begin{figure*}
\begin{center}
    \includegraphics{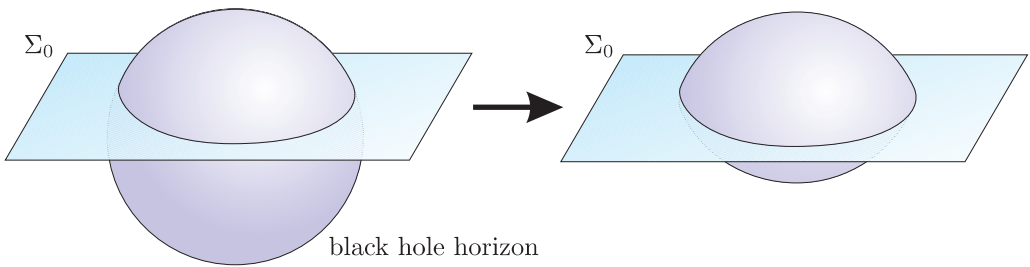}
\end{center}
\caption{An example of the displace--cut--reflect procedure when
the $\Sigma_0$ hypersurface is planar and intersects the horizon
of the bulk black hole.  In the picture, all but the $r$, $w$ and
one of the angular coordinates in the $S^{d-1}$ part of the metric
(\ref{cylindrical}) have been suppressed. Note that in this case,
we have elected to retain the singularity-free half of the
5-manifold in the braneworld model.}\label{fig:slicing}
\end{figure*}

Before moving on to the particular cases, we comment on the
general algorithm that we will employ.  The basic strategy for the
construction of braneworlds from bulk manifolds covered by
isotropic coordinates is the same as the 4-dimensional
``displace--cut--reflect'' procedure for constructing thin-disk
solutions to the Einstein equations \cite{Vogt:2003hi}. The key is
expressing the isotopic line element (\ref{isotropic line
element}) as
\begin{equation}\label{cylindrical}
    \ds{M} = -H(\rho)\,dt^2 + G(\rho) [dr^2 + r^2 \,
    d\Omega_{d-1}^2 + dw^2],
\end{equation}
where $\rho = \sqrt{r^2 + w^2}$.  This is nothing more than a
generalization of cylindrical coordinates on the isotropic spatial
section of (\ref{isotropic line element}).  To generate a
braneworld model, we pick one of the $w =$ constant hypersurfaces
$\Sigma_0$ to be the brane.  Naturally, the $\Sigma_0$
hypersurface will divide the bulk into two regions, one of which
we discard and replace with the mirror image of the other half. In
this way, we generate a $\mathbb{Z}_2$ symmetric braneworld model.
The metric on the brane is
\begin{eqnarray}\nonumber
    \ds{\Sigma_0} & = & -H \left( \sqrt{r^2 + w_0^2} \right) \,dt^2 + G \left( \sqrt{r^2 + w_0^2}
    \right) \\ & & \times [dr^2 + r^2 \, d\Omega_{d-1}^2],
    \label{general brane line element}
\end{eqnarray}
where $w = w_0$ is the defining equation of $\Sigma_0$.  We see
that the brane's geometry will necessarily be static and
spherically symmetric.  This procedure is diagrammed in Figure
\ref{fig:slicing}, where we show the case of a planar braneworld
intersecting a bulk black hole horizon.

One of the features of this procedure is that we have no control
over the extrinsic curvature of the $\Sigma_0$ hypersurface; it is
essentially fixed by the bulk geometry and our choice of a planar
braneworld geometry. Now, recall that in the general relativistic
thin-shell formalism, the matter carried by a geometric defect
such as $\Sigma_0$ is related in a direct way to its extrinsic
curvature. Therefore, the matter content of our braneworld is
given to us from the model, rather than being something that we
have input directly into the formalism.  What exactly is the
nature of the matter confined to $\Sigma_0$?  To answer this, we
need the normal to the family of $w =$ constant hypersurfaces
$\Sigma_w$:
\begin{equation}
    n_a = G^{1/2} \left( \sqrt{r^2 + w^2}
    \right) \di_a w.
\end{equation}
We need the projection tensor and extrinsic curvature associated
with $\Sigma_w$
\begin{equation}
    h_{ab} = g_{ab} - n_a n_b, \quad K_{ab} = h^{c}{}_{a} \nabla_c
    n_b,
\end{equation}
which leads to the following expression for the stress energy
tensor of matter on the brane:
\begin{equation}\label{brane stress-energy}
    \kappa_5^2 S_{ab} = -2(K_{ab} - h_{ab} \mathrm{Tr}K).
\end{equation}
Here, evaluation at $w = w_0$ is understood and $\kappa_5^2$ is
the 5-dimensional gravity-matter coupling.  In these coordinates,
we expect that $S_{ww} = 0$ since $S_{ab} n^a = 0$.  We will use
this expression below to read off the properties of the brane
matter in specific models.

\subsection{Braneworlds from a Schwarzschild bulk}\label{sec:branes from vacuum}

We now apply our braneworld construction to the isotropic
representation of the 5-dimensional Schwarzschild metric derived
in Section \ref{sec:iso Schwarzschild}.  Using equations
(\ref{isotropic Schwarzschild}) and (\ref{general brane line
element}), we obtain the following 4-geometry on the brane:
\begin{equation}\label{black shell metric}
    \ds{\Sigma_0} = -\left( \frac{r^2 + w_0^2 -1}{r^2 + w_0^2 +1}
    \right)^2 dt^2 + \left[ \frac{r^2 + w_0^2 + 1}{2(r^2+w_0^2)}
    \right]^2 d\sigma_3^2,
\end{equation}
where $d\sigma_3^2 = dr^2 + r^2 d\Omega_2^2$ is the metric on flat
Euclidean 3-space.  We will denote the metric on $\Sigma_0$ as
$h_{\alpha\beta}$.  This metric is static, spherically symmetric,
and asymptotically flat in the $r \rightarrow \infty$ limit. One
of the first things that one notices about this metric is that if
$w_0 \in [-1,1]$ there is a Killing horizon at $r = \sqrt{1-w_0^2}
\equiv r_0$ --- which we denote by $\mathcal{H}$
--- where the norm of $\di_t$ vanishes. It should be clear that
$\mathcal{H}$ is the intersection of the braneworld $\Sigma_0$
with the 5-dimensional black hole horizon $\rho = 1$. Now, it is
clear that the induced metric on $\mathcal{H}$ is degenerate with
signature $(0+++)$, hence it is a null surface as must be true for
all Killing horizons.

Another important feature of $\mathcal{H}$ is that it is the
location of a curvature singularity. To see this, consider the
Kretschmann curvature scalar:
\begin{equation}
    \mathcal{K} \equiv R^{\alpha\beta\gamma\delta} R_{\alpha\beta\gamma\delta} =
    \frac{1024 P(r,w_0)}{(r^2+w_0^2+1)^8 (r^2 + w_0^2 -1)^2},
\end{equation}
where $P(r,w_0)$ is a complicated 12$^\mathrm{th}$ order
polynomial in $r$ and $w_0$ satisfying $P(\sqrt{1-w_0^2},w_0) = 12
w_0^4$.  Therefore, $K$ diverges on $\mathcal{H}$ signifying that
the latter is a singular hypersurface.\footnote{For the moment, we
exclude the $w_0 = 0$ case.} Hence, in this spacetime a Killing
horizon and a curvature singularity are coincident. This is an
unusual, but not entirely unprecedented feature of this model. For
example, one sees similar behaviour in the extremal D$p$-brane
solutions of supergravity theory \cite{Peet:2000hn}. However, in
the majority of those geometries the Killing horizon is also an
event horizon. Is it the same true for this braneworld spacetime?

The answer is no, as we now demonstrate.  The key is to show that
there is a null geodesic of finite affine length that connects the
singularity with arbitrary points in the exterior region.
Radially outgoing geodesics in this spacetime have the following
tangent vector field in the affine parametrization:
\begin{equation}
    k^\alpha \di_\alpha = E \left( \frac{1}{H} \frac{\di}{\di t} +
    \frac{1}{\sqrt{GH}} \frac{\di}{\di r} \right),  \quad k^\alpha
    \nabla_\alpha k^\beta = 0,
\end{equation}
where $E$ is the energy parameter.  From this, we see that the
affine length $\Delta\lambda$ of a light ray travelling from
$\mathcal{H}$ to some $r_1 > r_0$ is
\begin{equation}
    \Delta \lambda = \frac{1}{E} \int_{r_0}^{r_1} \sqrt{GH} \, dr.
\end{equation}
The integrand here is manifestly finite, hence $\Delta \lambda$ is
similarly finite and $\mathcal{H}$ cannot be an event horizon.

Furthermore, $\mathcal{H}$ is not even a trapping
horizon.\footnote{Defined as the world tube of a series of
apparent horizons as in Ref.~\cite{Dreyer:2002mx}, for example.}
To see this, we introduce the time and radial unit vectors:
\begin{equation}
    \hat{t}^\alpha \di_\alpha = H^{-1/2}\,\di_t, \quad \hat{r}^\alpha
    \di_\alpha = G^{-1/2} \di_r.
\end{equation}
Then, for every 2-sphere $(t,r)=$ constant we can define vectors
tangent to ingoing and outgoing radial null congruences as
\begin{equation}
    \ell^\alpha = \frac{1}{\sqrt{2}} ( \hat{t}^\alpha -
    \hat{r}^\alpha ), \quad \tilde{k}^\alpha = \frac{1}{\sqrt{2}} ( \hat{t}^\alpha
    + \hat{r}^\alpha ),
\end{equation}
respectively.  Now, the induced metric on the 2-spheres is
$q_{\alpha\beta} = h_{\alpha\beta} + \hat{t}_\alpha \hat{t}_\beta
- \hat{r}_\alpha \hat{r}_\beta$ and the expansion of the ingoing
and outgoing congruences are
\begin{equation}
    \theta_{(\ell)} = q^{\alpha\beta} \nabla_\alpha \ell_\beta,
    \quad \theta_{(\tilde{k})} = q^{\alpha\beta} \nabla_\alpha \tilde{k}_\beta,
\end{equation}
respectively.  Now, we want to know whether or not the 2-spheres
that are the constant time slices of $\mathcal{H}$ are apparent
horizons.  They will be if the outgoing expansion scalar vanishes
for $r=r_0$.  A quick calculation shows:
\begin{equation}\label{apparent horizon}
    \theta_{(\tilde{k})} = \frac{r G_{,r} + 2  G}{2 r
    G^{3/2}},
\end{equation}
where $G_{,r} = dG/dr$.  It is straightforward to verify that this
reduces to
\begin{equation}
    \theta_{(\tilde{k})} = \frac{w_0^2}{\sqrt{1-w_0^2}},
\end{equation}
on $\mathcal{H}$.  Since the expansion is clearly non-zero for
$w_0 \ne 0$, we can conclude that $\mathcal{H}$ is not a trapping
horizon for such cases.

All this goes to show that when $w_0^2 \in (0,1]$, we are dealing
with a naked null singularity in this spacetime.  Interestingly,
we can find a coordinate system that is regular there.  More
precisely, the transformation
\begin{eqnarray}\nonumber
    u & = & -e^{-\tfrac{1}{2} r_0 (t - r_*)}, \quad v = e^{\tfrac{1}{2} r_0 (t +
    r_*)}, \\ r_* & = & \frac{r}{2} - \frac{1}{2w_0} \mathrm{arctan}
    \frac{r}{w_0} + \frac{1}{r_0} \ln \frac{r-r_0}{r+r_0},
    \label{Kruskal coordinates}
\end{eqnarray}
puts our metric in the form
\begin{eqnarray}\nonumber
    \ds{\Sigma_0} & = & -\frac{4(r+r_0)^4 \exp{\left(\frac{r_0}{2w_0} \mathrm{arctan}
    \frac{r}{w_0}-\frac{r_0r}{2}\right)}}{r_0^2(r^2-r_0^2+2)^2 }
     du\,dv \\ & & + \left[ \frac{r(r^2 + w_0^2 + 1)}{2(r^2+w_0^2)}
    \right]^2 d\Omega_2^2, \quad r = r(u,v).
\end{eqnarray}
All the metric coefficients are well behaved at $\mathcal{H}$,
despite the fact that there is a curvature singularity
there.\footnote{A singularity associated with a regular metric in
null coordinates is termed `weak' in the Tipler sense
\cite{Tipler,Burko:1997zy}.}

The final issue we want to address is the type of brane matter
that sources this model.  We can calculate the stress-energy
tensor for the $\Sigma_0$ hypersurface from the definitions
leading up to equation (\ref{brane stress-energy}); the result is:
\begin{equation}
    S^a{}_b = \mathrm{diag} \left(
    -\epsilon, p, p, p, 0
    \right),
\end{equation}
where
\begin{subequations}\label{density and pressure 1}
\begin{eqnarray}
    \kappa_5^{2} \epsilon & = & \frac{24w_0}{(r^2 + w_0^2+1)^2},
    \\ \kappa_5^{2} p & = & \frac{16w_0}{(r^2 + w_0^2+1)^2 (r^2 +
    w_0^2-1)}.
\end{eqnarray}
\end{subequations}
Therefore the brane matter admits a perfect fluid type description
with energy density $\epsilon$ and isotropic pressure $p$.  For
$w_0 > 0$, the density is finite and positive for all $r$.  On the
other hand, the pressure is changes sign from positive to negative
and diverges as $r$ decreases across $r=r_0$.  So, in addition to
a curvature singularity at $r = r_0$, we also have a singularity
in some of the matter properties.\footnote{As an interesting
aside, we note that one can also get pressure singularities when a
`bouncing' cosmological brane is embedded in a Schwarzschild bulk
\cite{PonceDeLeon:2001un,Liu:2002pk,Xu:2003fb,Wang:2003yr}. In
that case, the origin of the singularity is a cusp in the
embedding functions at the position of the bounce, which can occur
within the bulk black hole event horizon \cite{Seahra:2003bu}.}

There are two comments to be made about the brane matter:  The
first centers around the observation that for $w_0 < 0$, the
exterior density and pressure are both negative. The reason for
this comes from an implicit assumption in our derivation; namely,
we always discard the part of the bulk manifold with $w < w_0$
when constructing our braneworld.  If $w_0 < 0$, then there will
be a 5-dimensional black hole on either side of the brane in
static equilibrium. The only way to keep the black holes from
crashing into each other is to separate them with a concentration
of repulsive matter; i.e., matter with $\epsilon + 3p < 0$.  Hence
the negative energy when $w_0 < 0$.

Our second comment has to do with the pressure singularity on
$\mathcal{H}$.  In 5 dimensions, the brane can be thought of as a
static thin disk of matter, and the disk's pressure provides
support against gravitational collapse.  But recall that an
infinite amount of force is required to maintain a static matter
distribution infinitessimally close to the surface of a black
hole. Since $\mathcal{H}$ is the intersection of the brane with
the 5-dimensional horizon, we see that the pressure singularity is
needed to prevent the disk matter from falling into the black
hole.  As viewed from the brane, we have a spherical distribution
of matter on the verge of gravitational collapse supported by a
shell-like pressure singularity.

To summarize, we have employed a planar slicing of the isotropic
coordinate patch of the Schwarzschild 5-manifold derived in
Section \ref{sec:iso Schwarzschild} to derive a class of
braneworld models (\ref{black shell metric}).  The models are
static and spherically symmetric, and characterized by a singular
null Killing horizon.  We also derived the properties of the brane
matter supporting the 4-geometry, for which there is an effective
perfect fluid description.  The energy density is well-behaved,
but we found that the pressure had singular behaviour on the
Killing horizon $\mathcal{H}$.  The pressure singularity is needed
to prevent the collapse of the brane into the 5-dimensional event
horizon.

\subsection{Braneworlds from a Schwarzschild-AdS bulk}

We now move on to the case of Schwarzschild-AdS bulk manifolds.
While the individual calculations are somewhat more involved than
those of the previous section, the procedures and results are
fairly similar.  In this case the brane metric is:
\begin{subequations}\label{AdS brane metric}
\begin{eqnarray}
    \ds{\Sigma_0} & = & -H\,dt^2+G\,(dr^2+r^2\,d\Omega_2^2),\\
    G & = & \frac{ \nc^2\left( \varphi , s_\gamma
    \right) }{r^2+w_0^2}, \\
    H & = & \frac{ \Jacobisc^2 \left(
    \varphi , s_\gamma \right) + a_\gamma^2 \sn^2
    \left( \varphi , s_\gamma \right)  } {\gamma^2}, \\
    \varphi & = & \tfrac{1}{2}\xi_\gamma  \ln(r^2+w_0^2).
\end{eqnarray}
\end{subequations}
As in the last section, we will suppress the $\sqrt{r^2+w_0^2}$
argument of the various metric functions.  It is useful to have
series expansions of $G$ and $H$ about $r = r_0 \equiv
\sqrt{1-w_0^2}$, which are:
\begin{subequations}
\begin{eqnarray}\nonumber
    G & = & 1 - 2r_0(r-r_0)+ [r_0^2(\xi_\gamma^2+4)-1](r-r_0)^2 +
    \\ & & \mathcal{O}[(r-r_0)^3], \\
    H & = & \frac{\xi_\gamma^2 r_0^2}{\gamma^2} (a_\gamma^2 +1) (r - r_0)^2 +
    \mathcal{O}[(r-r_0)^3].
\end{eqnarray}
\end{subequations}
It is immediately obvious from these series that the $r = r_0$
hypersurface is again a Killing horizon $\mathcal{H}$.  Also,
since we can directly apply equation (\ref{apparent horizon}) to
this situation, we can use the above series expansion for $G$ to
obtain the expansion of an outgoing null congruence on
$\mathcal{H}$:
\begin{equation}
    \theta_{(\tilde{k})} = \frac{w_0^2}{\sqrt{1-w_0^2}}.
\end{equation}
This is precisely the same result as in the vacuum bulk case and
leads us to the same conclusion: $\mathcal{H}$ is neither an event
nor trapping horizon.

To determine if $\mathcal{H}$ is the location of a curvature
singularity, we can use the exact expressions for $G$ and $H$ to
calculate the Kretschmann scalar, and then perform another
expansion about $r = r_0$.  The result is:
\begin{equation}
    K = \frac{12(1-r_0)^2(1+r_0)^2}{r_0^2} (r-r_0)^{-2}
    +\mathcal{O}[(r-r_0)^{-1}].
\end{equation}
This clearly diverges as $r \rightarrow r_0$, so we have that
$\mathcal{H}$ is the site of a curvature singularity.
Furthermore, when this fact is coupled with our knowledge of the
fact that $\mathcal{H}$ is not an event horizon, we conclude that
it is a naked singularity, just as before.

%We can address the issue of whether or not is a weak singularity
%in a similar manner as before, but the complexity of the metric
%functions requires that we work very close to $\mathcal{H}$.
%Recall, that the goal is to find a coordinate transformation that
%puts the metric into a well-behaved form near the singularity.
%Such a transformation is given by
%\begin{subequations}\label{AdS Kruskal transformation}
%\begin{eqnarray}
%    r_* & = & \int \sqrt{\frac{G}{H}} \, dr \approx
%    \frac{\gamma^2 \ln(r -r_0) }{r_0(\gamma^2+2)}, \\
%    u & = & -\exp \left[ - \frac{r_0(\gamma^2+2)}{\gamma^2}(t-r_*) \right], \\
%    v & = & +\exp \left[ + \frac{r_0(\gamma^2+2)}{\gamma^2}(t+r_*)
%    \right],
%\end{eqnarray}
%\end{subequations}
%which puts the metric in the form
%\begin{equation}
%    \ds{\Sigma_0} \approx - du\,dv +r^2 d\Omega_2^2,
%\end{equation}
%near the singularity.  This is obviously well-behaved, so we
%conclude that the singularity is weak in the Tipler sense.

\begin{figure}
\begin{center}
\includegraphics{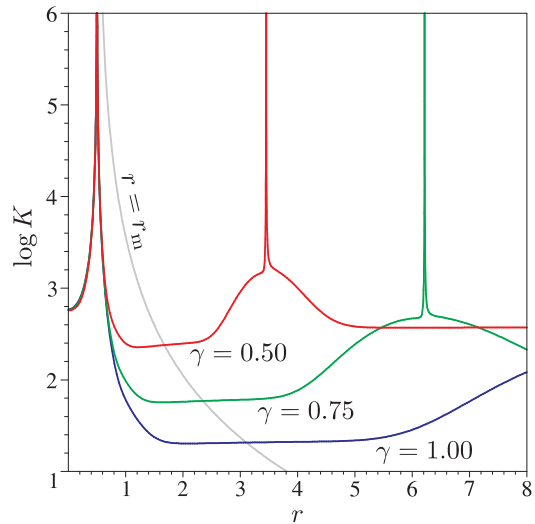}
\end{center}
\caption{The Kretschmann scalar as a function of $r$ for various
braneworlds in the case of an S-AdS$_5$ bulk.  Each of the sharp
vertical peaks represent infinite spikes.  We have selected
$w_0=\sqrt{3}/2$ and as expected, we see the divergence of $K$ at
$r = r_0 = 1/2$ in all instances. The intersection of the $r =
r_\mathrm{m}$ line with each of the other curves gives the
position of spatial infinity for each value of $\gamma$.  Hence,
we see that the rightmost spikes are `beyond infinity'; i.e.,
there is only one divergence of $\log K$ for $r \in
[r_0,r_\mathrm{m})$.}\label{fig:K}
\end{figure}

One distinctive feature of this case is the asymptotic structure.
Recall that when we derived the isotropic patch for S-AdS$_5$, the
entirety of the region outside the black hole was covered by a
finite interval of isotropic radius $\rho \in
(1,\rho_\mathrm{max})$.  This would lead us to expect that there
might be some special behaviour of the 4-dimensional model at
$r_\mathrm{m} = \sqrt{\rho_\mathrm{max}^2-w_0^2}$.  Now, our
previous formula for $\rho_\mathrm{max}$ (\ref{max rho}) gives us
that $\varphi = K(s_\gamma)$ at $r = r_\mathrm{m}$, which allows
us to expand our metric functions about $r = r_\mathrm{m}$.
Keeping leading order terms only, we have:
\begin{equation}\label{asymptotic metric}
    \ds{\Sigma_0} \sim
    \frac{r_\mathrm{m}^2+w_0^2}{r_\mathrm{m}^2 (r
    -r_\mathrm{m})^2}  [-dt^2+\gamma^2 (dr^2 + r_\mathrm{m}^2 d\Omega_2^2)
    ].
\end{equation}
From this, it is clear that the proper distance between any point
with $r \in (r_0,r_\mathrm{m})$ and the $r = r_\mathrm{m}$
hypersurface is infinite.  Hence, we should regard $r =
r_\mathrm{m}$ as the spatial infinity of our 4-geometry.  With
this understanding, we can now interpret the plots of the brane's
Kretschmann scalar versus $r$ shown in Figure \ref{fig:K}.  These
show the expected divergence of $K$ at $r = r_0$, but there are
additional infinite features at greater values of $r$.  As
explained in the caption, these spikes always occur at $r >
r_\mathrm{m}$ and hence are `beyond infinity'; hence, they need
not overly concern us.

We now determine the asymptotic behaviour of the geometry as $r
\rightarrow r_\mathrm{m}$ by calculating the limiting value of the
Riemann tensor.  To lowest order in $(r-r_\mathrm{m})$, we find:
\begin{equation}
    R^{\alpha\beta}{}_{\gamma\delta} = - \frac{r_\mathrm{m}^2}{\gamma^2 (r_\mathrm{m}^2+w_0^2)
    } (\delta^{\alpha}{}_{\gamma}
    \delta^\beta{}_\delta - \delta^\beta{}_\gamma \delta^\alpha{}_\delta
    ).
\end{equation}
Hence, we have an asymptotically AdS-structure for the 4-geometry
with total cosmological constant:\footnote{This was foreshadowed
by the form of the asymptotic metric (\ref{asymptotic metric}),
which suggests that null geodesics could travel an infinite proper
distance in a finite amount of coordinate time --- a hallmark of
AdS-space.}
\begin{equation}
    \Lambda_4 = - \frac{3r_\mathrm{m}^2}{\gamma^2 (r_\mathrm{m}^2+w_0^2)
    }.
\end{equation}
This can be compared with the 5-dimensional cosmological constant
sourcing the bulk:
\begin{equation}
    \Lambda_5 = -\frac{6}{\gamma^2}.
\end{equation}
In situations such as these, there are standard formulae that
relate $\Lambda_5$ and $\Lambda_4$ with the brane's tension
$\lambda$ (see Ref.~\cite{Maartens:2003tw}, for example).  In
particular:
\begin{equation}\label{tension 1}
    \lambda = \pm \sqrt{ 6(2\Lambda_4-\Lambda_5) } = \pm \frac{6}{\gamma} \sqrt{
    \frac{w_0^2}{r_\mathrm{m}^2+w_0^2}}.
\end{equation}
To remove the sign ambiguity in the sign of $\lambda$, we need to
look at the properties of the brane matter sourcing the model,
which are obtained using the general algorithm outlined above. The
resulting expression for the brane's stress energy tensor is
extremely complicated and writing it down here will not convey
much insight.  But there are a few points worth mentioning:
\begin{itemize}

\item The brane stress-energy tensor is of the perfect fluid type
with $S^a{}_b = \mathrm{diag}(-\epsilon,p,p,p,0)$, just as for the
vacuum-bulk case.

\item At $r = r_0$, the density and pressure behave like:
\begin{equation}
    \lim_{r \rightarrow r_0} \kappa_5^2 \epsilon = 6 w_0,  \quad
    \lim_{r \rightarrow r_0} \kappa_5^2 p = \frac{2 w_0}{r_0 (r -
    r_0)};
\end{equation}
i.e., the density is finite at the position of the curvature
singularity, while the pressure has an asymmetric pole such that
$p > 0$ for $r>r_0$.  This exactly mirrors the vacuum-bulk case
[cf.~equations (\ref{density and pressure 1})].

\item We have the following limiting behaviour near spatial
infinity:
\begin{subequations}
\begin{eqnarray}
    \lim_{r \rightarrow r_\mathrm{m}^-} \kappa_5^2 \epsilon & = & -
    \frac{6w_0}{\gamma\sqrt{r_\mathrm{m}^2+w_0^2}}, \\
    \lim_{r \rightarrow r_\mathrm{m}^-} \kappa_5^2 p
    & = & + \frac{6w_0}{\gamma\sqrt{r_\mathrm{m}^2+w_0^2}}.
\end{eqnarray}
\end{subequations}
Therefore, the asymptotic behaviour of the brane matter is that of
vacuum energy with cosmological constant
\begin{equation}
    \lambda = -\frac{6w_0}{\gamma\sqrt{r_\mathrm{m}^2+w_0^2}}.
\end{equation}
We identify this as the tension of our brane, which is
\emph{negative} for $w_0 > 0$. Of course it is in complete
agreement with equation (\ref{tension 1}), which was obtained from
a direct analysis of the asymptotic geometry.

\end{itemize}
All of these features are manifest in Figure \ref{fig:density
pressure}, which shows $\epsilon$ and $p$ for a particular choice
of $\gamma$ and $w_0$.
\begin{figure}
\begin{center}
\includegraphics{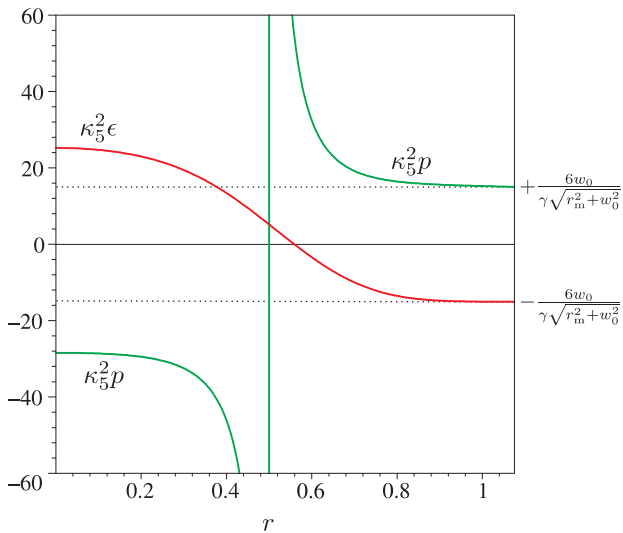}
\end{center}
\caption{Density and pressure of brane matter for $\gamma = 1/4$
and $w_0 = \sqrt{3}/2$.  Note for these parameters $r_\mathrm{m}
\approx 1.08$, which has been selected as the rightmost point on
the $r$-axis.}\label{fig:density pressure}
\end{figure}

To summarize this section, we have used planar slicings of the
isotropic coordinate map of S-AdS$_5$ to generate braneworld
models.  The 4-manifolds we obtained share many properties with
the ones derived from purely vacuum 5-manifolds in \ref{sec:branes
from vacuum}; in particular they involve null, shell-like, naked
singularities where the brane crosses the bulk black hole horizon.
The models approach AdS$_4$ in the (appropriately identified)
asymptotic region, and they are supported by perfect fluid brane
matter.  The pressure diverges at the position of the singularity
while the density remains finite.  By calculating the asymptotic
behaviour of $\epsilon$ and $p$, we found an explicit expression
for the brane tension, which is negative if $w_0 > 0$.

\section{Braneworlds from non-planar
slicings}\label{sec:non-planar}

In the previous section, we saw that our consideration of purely
planar slicings of black hole 5-manifolds led to brane matter
whose properties were largely given to us by the geometry.  In
order to regain some control of the sources in our model, we now
consider braneworlds formed from surfaces of revolution. For
simplicity, we will limit our work to bulk vacuum bulk manifolds,
though much of what we do can be straightforwardly generalized to
the S-AdS$_5$ case.

We can define a surface of revolution in isotropic coordinates by
$w = w(r)$, which induces the following metric on $\Sigma_0$:
\begin{equation}
    \ds{\Sigma_0} = -H \,dt^2 + G [ (1+w_{,r}^2)\,dr^2 +
    r^2\,d\Omega_2^2 ],
\end{equation}
where $w_{,r} = dw/dr$ and $H$ and $G$ are the isotropic metric
functions (\ref{isotropic Schwarzschild}) evaluated at $\rho =
\sqrt{r^2 + w^2(r)}$.  Like the planar case, the 4-geometry is
static and spherically symmetric.  Now, let us calculate the Ricci
scalar for this geometry:
\begin{equation}
    {}^{(4)}R = -\frac{8 q_1}{r^2(r^2 + w^2 + 1)^3 (r^2+w^2-1)
    (1 + w_{,r}^2)^2},
\end{equation}
where $q_1 = q_1(r,w,w_{,r},w_{,rr})$ is given in the Appendix and
we have written $w_{,rr} = d^2 w/dr^2$. For generic choices of
$w(r)$, the Ricci scalar will diverge at $r = r_0$, where $r_0$ is
the solution of $r^2 + w^2(r) = 1$.\footnote{However, there is one
important special case we should highlight: namely $w(r) =
\text{constant} \times r$, which is equivalent to $\chi = \chi_0$
in the original Schwarzschild coordinates. In this case, we find
$q_1$ vanishes identically; i.e., ${}^{(4)}R = 0$. Indeed, the
complete set of 4-dimensional curvature invariants is regular at
$r = r_0$, so this braneworld likely does not have a shell
singularity.  I would like to thank Ken-ichi Nakao and Daisuke Ida
for drawing this case to my attentiion.} We therefore identify a
curvature singularity at the $r=r_0$ hypersurface, just as in the
planar case.  We again expect this singularity to be naked,
because in this case the affine length of a radial null geodesic
with energy $E$ travelling from $r=r_1$ to $r_0$ is
\begin{equation}
    \Delta \lambda = \frac{1}{E} \int_{r_0}^{r_1} \sqrt{GH(1+w_{,r}^2)} \,
    dr,
\end{equation}
which is generally finite, except perhaps for very special choices
of $w(r)$.

Hence, the non-planar case is similar to the planar case in that
we can expect to find naked shell singularities.  However, we now
have an additional degree of freedom at our disposal to place
certain conditions on the geometry, the brane matter, or both. Let
us now consider a few examples of how this freedom can be used.

\subsection{Slicings with vanishing Ricci scalar}

\begin{figure}
\begin{center}
\includegraphics{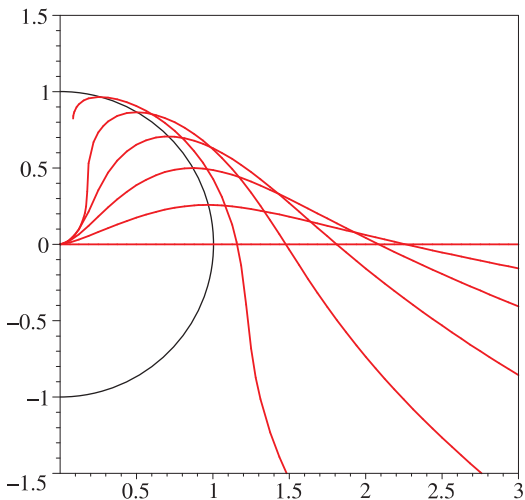}
\end{center}
\caption{Numeric solutions for $w(r)$ associated with ${}^{(4)}R =
0$ slicings. The semi-circle indicates the position of the
horizon. Note that the topmost curve is incomplete because it has
a vertical tangent when inside the horizon, which caused the
numeric integration to fail.}\label{fig:ricci flat}
\end{figure}

In the, admittedly short, history of the search for braneworld
black holes, many workers have viewed solutions of ${}^{(4)}R = 0$
as likely candidates.  This is because of one of the contracted
Gauss-Codazzi equations, which for a vacuum bulk reads:
\begin{equation}
    {}^{(4)}R = (\mathrm{Tr}K)^2 - K^{ab} K_{ab}.
\end{equation}
Since the brane's stress-energy tensor is essentially determined
by the extrinsic curvature, ${}^{(4)}R$ must vanish for models
with no brane matter present.  However, the reverse is not true;
if ${}^{(4)}R=0$ we do not necessarily have that $K_{\alpha\beta}
= 0$. In our case, the imposition of ${}^{(4)}R = 0$ will not
guarantee that the extrinsic curvature of the brane vanishes ---
it is actually impossible to get $K_{\alpha\beta} = 0$ for a
non-trivial slice (\emph{cf.} Sec.~\ref{sec:zero extrinsic}).
However, it is still an interesting case to look at because it
does place a constraint on the total effective matter on the
brane, which includes contributions from both the brane and `Weyl'
matter. In order to obtain $w(r)$ we set $q_1 = 0$ and solve the
resulting second-order ODE numerically. Several representative
solutions for $w(r)$ are plotted in Figure \ref{fig:ricci flat}.
In this plot, we note the planar solution $w = 0$ along with more
exotically shaped braneworlds.

\subsection{Slicings with vanishing radial pressure}

\begin{figure}
\begin{center}
\includegraphics{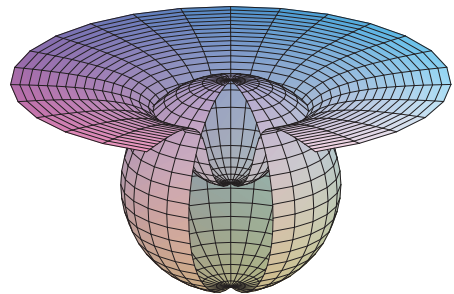}
\end{center}
\caption{The surface of revolution formed from our analytic
solution for $w(r)$ in the case of a $p_r = 0$ slicing (a wedge
has been removed to aid visualization).  The spherical object
indicates the position of the black hole horizon. The time and
$\theta$ coordinates have been suppressed, which means that each
horizontal ruling on the surfaces actually represents a
2-sphere.}\label{fig:radial pressure}
\end{figure}

Another class of interesting braneworld are those with vanishing
principle pressures.  Using equation (\ref{brane stress-energy})
with an extrinsic curvature calculated from
\begin{equation}
n_a dx^a = \frac{-w_{,r} dr + dw}{\sqrt{(w_{,r}^2+1)G}},
\end{equation}
we find that the brane's stress-energy tensor is of the form
\begin{equation}
     S^a{}_b = \mathrm{diag} (-\epsilon,p_r,p_\bot,p_\bot,0),
\end{equation}
where $p_r \ne p_\bot$ in general.  Explicitly, the radial
pressure is
\begin{equation}
    \kappa_5^2 p_r = \frac{q_2}{r(r^2+w^2+1)^2(r^2+w^2-1)\sqrt{1+w_{,r}^2}},
\end{equation}
where $q_2 = q_2(r,w,w_{,r})$ is given in the Appendix.  In this
expression, we see the now familiar pressure singularity at $r^2 +
w^2(r) = 1$. To find a braneworld with zero radial pressure, we
need to solve the first-order ODE $q_2 = 0$.  This is actually
possible to do in a closed form, and the exact solution is given
in the Appendix. We give a 3-dimensional representation of one
possible braneworld obtained from this solution in Figure
\ref{fig:radial pressure}. The plot gives the impression that the
brane approaches a planar geometry from large $r$, which is
actually not true since the limit of $|w(r)|$ as $r \rightarrow
\infty$ is itself infinite.

\subsection{Slicings with vanishing tangential pressure}

\begin{figure}
\begin{center}
\includegraphics{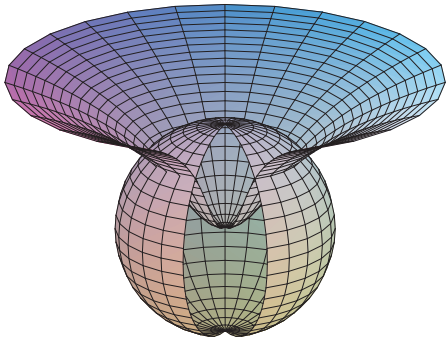}
\end{center}
\caption{A numeric solution for $w(r)$ in the case of a $p_\bot =
0$ slicing (see the caption of Figure \ref{fig:radial pressure}
for an explanation of the visualization
scheme)}\label{fig:tangential pressure}
\end{figure}

We now turn our attention to braneworlds where the brane matter
satisfies $p_\bot = 0$.  In general, we have
\begin{equation}
    \kappa_5^2 p_\bot =
    \frac{-4q_3}{r(r^2+w^2+1)^2(r^2+w^2-1)(1+w_{,r}^2)^{3/2}},
\end{equation}
where $q_3 = q_3(r,w,w_{,r},w_{,rr})$ is given in the Appendix.
Obviously, the ODE to solve for $p_\bot = 0$ is $q_3 = 0$, which
is a rather complicated expression.  We will content ourselves
with numerical solutions, one of which is depicted in Figure
\ref{fig:tangential pressure}.

\subsection{Slicings with isotropic pressure}

We have already seen above that the planar slicings of
5-dimensional Schwarzschild space give rise to braneworlds with
$p_r = p_\bot$; i.e., with isotropic pressure.  But are planar
slicings the only ones that can be modelled as a perfect fluid? To
answer this, consider:
\begin{equation}
    \kappa_5^2 (p_r - p_\bot) = \frac{4(r^2+w^2)(r w_{,rr}-w_{,r}-w_{,r}^3)}
    {(w_{,r}^2+1)^{3/2}(w^2+r^2+1)r}
\end{equation}
Setting this equal to zero, we find
\begin{equation}
    c_1^2 = r^2 + (w - c_2)^2,
\end{equation}
where $c_1$ and $c_2$ are arbitrary constants.  Hence braneworlds
with isotropic pressure have circular cross-sections in the
$(r,z)$-plane and look like off-center spheres as surfaces of
revolution.  In the limit of large radius ($c_1 \rightarrow
\infty$), we recover the planar result of Sec.~\ref{sec:branes
from vacuum}.

\subsection{Slicings with vanishing extrinsic
curvature}\label{sec:zero extrinsic}

We now turn our attention to braneworlds with $K_{ab} = 0$, which
represent models with no matter confined to $\Sigma_0$.  If we
calculate the extrinsic curvature explicitly, we find
\begin{equation}
    K^t{}_t = \frac{8(r^2+w^2)(r w_{,r}-w)}
    {\sqrt{w_{,r}^2+1}(1-w^2+r^2)(w^2+r^2+1)^2}.
\end{equation}
Setting this equal to zero yields $w = cr$, where $c$ is a
constant.  Plugging this into $K^a{}_b$, we find the only
non-vanishing components:
\begin{equation}
    K^\theta{}_\theta = K^\phi{}_\phi = -\frac{2cr\sqrt{c^2+1}}
    {(c^2+1)r^2+1}.
\end{equation}
Setting these identically equal to zero implies $c = 0$.  Hence
the only surface of revolution we can find with $K_{ab} = 0$ is $w
= 0$; i.e., the equatorial plane of the black hole.  This result
makes intuitive sense, because we know that surfaces with
vanishing extrinsic curvature must be symmetry surfaces of our
spacetime, and the only way to symmetrically slice our 5-manifold
is down the middle.

\subsection{Slicings resulting in a pure tension brane:
Einstein-static universe}\label{sec:Einstein-static}

The last type of slicing that we consider has the extrinsic
curvature proportional to the induced metric, which represents a
brane sourced by a cosmological constant (i.e.~tension) only. Such
a slicing has been previously sought by Chamblin et
al.~\cite{Chamblin:1999by}, but due to a particular choice of
embedding scheme was not found.\footnote{See also the work of
Kodama \cite{Kodama:2002kj,Kodama:2002pi}, which searched for pure
tension branes in quite general bulk manifolds satisfying minimal
symmetry assumptions.} Assuming $K_{ab} = -\tfrac{1}{6} \sigma
h_{ab}$ yields the solution:
\begin{equation}
    r^2 + w^2 = 3 + 2\sqrt{2}, \quad \sigma = \pm 3.
\end{equation}
Hence, the only pure tension brane solution takes the form of a
static spherical shell of isotropic radius $\rho = \sqrt{3 +
2\sqrt{2}}$, which corresponds to a Schwarzschild radius of $R =
\sqrt{2}$. The 4-metric in this case can be cast as
\begin{equation}
    \ds{\Sigma_0} = -d\tau^2 + 2 \,d\Omega_3^2,
\end{equation}
where $\tau = t/\sqrt{2}$.  This is the metric of the
Einstein-static universe.  In other words, pure tension static
branes around 5-dimensional Schwarzschild black holes take the
form of an Einstein-static universe.  Note that the `dark
radiation' plays the role that matter would in a 4-dimensional
Einstein static solution, as can be seen from the effective
Friedman and Raychaudhuri equations of this `brane cosmology',
which in dimensionless coordinates read:
\begin{equation}
    \frac{1}{R^2} \left( \frac{dR}{dt} \right)^2 = -\frac{1}{R^2}
    + \frac{1}{R^4} + \frac{\sigma^2}{36} \equiv 0, \quad \frac{d^2 R}{dt^2} \equiv 0.
\end{equation}

An interesting observation is that in this case, the brane is
coincident with the photon sphere of the bulk black hole.  This
fact could have been anticipated from the following fact: Any
5-dimensional null geodesic initially tangent to a pure tension
brane $\Sigma_0$ will remained confined to that brane. This can be
seen by noting the following result from
Ref.~\cite{Seahra:2003kc}: Given that a null geodesic is
momentarily tangent to a hypersurface $\Sigma_0$, its acceleration
orthogonal to that surface is proportional to $K_{ab} k^a k^b$,
where $k^a$ is the tangent vector.  If $\Sigma_0$ is a pure
tension brane we have $K_{ab} = -\tfrac{1}{6} \sigma h_{ab}$, from
which it follows $K_{ab} k^a k^b = 0$ since $k^a n_a = k^a k_a =
0$. Hence, there is no acceleration perpendicular to $\Sigma_0$
and null geodesics are confined to pure tension branes.  Surfaces
such as this are known as `totally geodesic' with respect to null
paths, which are a special type of umbilical surface
\cite{Kodama:2002kj}.

\section{Conclusions}\label{sec:conclusions}

In this paper we have considered braneworld models obtained by
non-trivial slicings of S-\ads{5} manifolds defined in isotropic
coordinates (Sec.~\ref{sec:isotropic}). We have succeeded in
finding a number of static and spherically symmetric
configurations, but almost all of them are characterized by a
naked pressure singularity where the brane crosses the horizon of
the bulk black hole.  From a relativity point of view, such a
singularity is required to provide the infinite force supporting
matter infinitesimally above an event horizon.  From the AdS/CFT
perspective, such a singularity can be interpreted as
Boulware-type quantum correction to the horizon of the brane black
hole.  Generic models have non-zero matter content; for planar
slicings we recover perfect fluid matter with an exotic equation
of state (Sec.~\ref{sec:branes from isotropic charts}).  Different
possible constraints on the 4-geometry were considered in
Sec.~\ref{sec:non-planar} in the simpler case of zero bulk
cosmological constant.  Branes with zero Ricci scalar, extrinsic
curvature, vanishing principle pressures, and others were derived;
but the only solution with vacuum (i.e., only brane tension)
turned out to be the braneworld generalization of the
Einstein-static universe residing on the 5-dimensional
photon-sphere.

It must be said that none of the braneworlds derived can be
considered as a black hole candidate.  The ubiquitous matter
content precludes that, but some interesting points have been
raised nevertheless.  We have explicitly seen how a regular bulk
can easily give rise to a singular brane, and how singular
4-dimensional horizons are a persistent feature of our
construction. Because this is from the divergence of tidal forces
on static matter near the surface of a black hole, we expect it to
generalize to any static brane with matter that intersects a bulk
Killing horizon.  Whether or not this extends to the `real'
\emph{static vacuum} braneworld black hole solution is an open
question: it is unclear if one needs a pressure singularity to
support the Weyl fluid certain to be present in such a model.  If
so, this provides strong support for, and physical insight into,
the conjecture that braneworld black holes naturally incorporate
quantum corrections.

One important issue that we have not addressed is the stability of
these models.  While it is true that the bulk geometries are
stable, there is no guarantee that the inclusion of a brane
boundary will not have a destabilization effect.  Actually finding
out if these models are stable is not an easy task, since all the
branes considered tend to break the $S^3$ symmetry of the bulk,
which complicates the analysis of perturbation wave equations. The
exception is the Einstein-static brane universe seen in
Sec.~\ref{sec:Einstein-static}, which is prone to a relatively
straightforward stability analysis.  We will report on this case
in a forthcoming paper.

\begin{acknowledgments}
I would like to thank Roy Maartens for interesting discussions and
encouragement, Ken-ichi Nakao for feedback, Mariam Bouhmadi for
advice on elliptic functions, and NSERC for financial support.
\end{acknowledgments}

\appendix

\begin{widetext}

\section*{Appendix}

\begin{itemize}

\item Definitions of various quantities associated with non-planar
slicings:
\begin{subequations}
\begin{eqnarray}
\nonumber q_1 & = & \left[(8 r^{7} w^{2} + 2 w^{8} r + 2 r^{9 } +
12 r^{5} w^{4} + 8 r^{3} w^{6} + 2 r^{5} - 2
w^{4} r) w_{,r} - 4 r^{2} w^{3} - 4 r^{4} w \right] w_{,rr} \\
\nonumber & & + (w^{8} - w^{4} + 6 w^{2} r ^{2} - 5 r^{4} + 6
r^{4} w^{4} + 4 r^{2} w^{6} + r^{8}
+ 4 r^{6} w^{2}) w_{,r}^4 + (16 w r^{3} - 8 r w^{3}) w_{,r}^3 \\
& & + ( - 6 w^{2} r^{2} - 5 r^{4} - w^{4} + 4 r^{6} w^{2} + r^{8}
+ 4 r^{2} w^{6}
 + w^{8} + 6 r^{4} w^{4})
w_{,r}^{2} + (16 w
 r^{3} - 8 r w^{3}) w_{,r} - 12 w^{2} r^{2},
 \\ q_2 & = & (-8 w^6-24 r^2 w^4+8 w^2-8 r^2-8 r^6-24 r^4 w^2) w_{,r} +16 w
 r, \\ \nonumber
 q_3 & = & [r^7+3 w^2 r^5+(3 w^4-1) r^3+(-w^2+w^6) r]
 w_{,rr} +[r^6+3 r^4 w^2+(3+3 w^4) r^2-w^2+w^6]
 w_{,r}^3 \\ & & -4 r w w_{,r}^2+[r^6+3 r^4 w^2+(3+3 w^4)
 r^2-w^2+w^6] w_{,r}-4 w r.
\end{eqnarray}
\end{subequations}

\item Analytic solution for $w(r)$ in the case of vanishing radial
pressure:
\begin{subequations}
\begin{eqnarray}
w(r) & = & \pm \frac{\sqrt{2^{7/3} X^4+8 c r^2 - 12 + 4 r^4 + 4
c^2 - 2^{14/3} r^2  X^2 + 2^{8/3} c X^2}}{864^{1/6} X}, \\
\nonumber X^6 & \equiv &  3 (24 r^4+12 r^8-60 c r^2+36 r^6 c+12
c^3 r^2+12-3 c^2+36 c^2 r^4)^{1/2} \\ & & 2 r^6+6 c r^4+(6 c^2+18)
r^2+2 c^3 -9 c,
\end{eqnarray}
\end{subequations}
where $c$ is an arbitrary constant.

\end{itemize}

\end{widetext}

\bibliography{black_shell}

\end{document}